# On Extraction of Chemical Potentials of Quarks from Particle Transverse Momentum Spectra in High Energy Collisions

Hong Zhao and Fu-Hu Liu[1]

*Institute of Theoretical Physics, Shanxi University, Taiyuan, Shanxi 030006, China*

**Abstract:**

We present two methods to extract the chemical potentials of quarks in high energy collisions. The first method is based on the ratios of negatively/positively charged particles, and the temperatures extracted from the transverse momentum spectra of related hadrons are needed. The second method is based on the chemical potentials of some particles, and we also need the transverse momentum spectra of related hadrons. To extract the quark chemical potentials, we would like to propose experimental collaborations to measure simultaneously not only the transverse momentum spectra of $\bar{p}$, $p$, $K^-$, $K^+$, $\pi^-$, and $\pi^+$, but also those of $D^-$, $D^+$, $B^-$, and $B^+$ (even those of $\Delta^{++}$, $\Delta^-$, and $\Omega^-$) in high energy nuclear collisions.

## 1. Introduction

The physics of high energy collisions is considered as an important frontier of fundamental researches which embrace fields such as particle physics, nuclear physics, and astrophysics [1–3]. A lot of final-state relativistic particles are produced and nuclear fragments are emitted in the collisions. Over an energy range from a few tens of MeV to a few TeV per nucleon, a large number of experimental data have been obtained from fixed target experiments to collider experiments, and many theoretical models have been introduced to describe the different experimental results [4–6]. The collisions within a very short period will not only present a colorful phenomenon, but also provide lots of information to comprehend the strong interacting theory and nuclear reaction mechanisms.

The constructions of the Relativistic Heavy Ion Collider (RHIC) and the Large Hadron Collider (LHC) have provided unprecedented opportunities to study nuclear and quark matters. The comprehension of the properties of nuclear matter at extreme conditions of high temperature and high density is crucial to search for the evidence of a new form of matter called Quark-Gluon Plasma (QGP) or quark matter [7–9], and to study the origin of the universe. Considerable attentions have been paid on the mechanisms of the formation and evolution of the QGP. It is generally believed that the QGP produced in the relativistic heavy ion collisions is in a thermal non-equilibrium state during initial stage and then evaluates into equilibrium state. However, how to describe the non-equilibrium state of QGP is still an unsolved and important problem. For any system, we can determine the direction and limitation of mass transfer by comparing the chemical potential of mater, that is to say the chemical potential is a sign to mark the direction of spontaneous chemical reaction. In fact, the chemical potential is also a criterion for determining whether thermodynamic equilibrium does exist in the interacting region in relativistic collisions.

---

[1] E-mail: fuhuliu@163.com; fuhuliu@sxu.edu.cn

Consequently, the chemical potential is one of the major problems for researching the QGP. In literature [10–12], different theoretical models have been proposed to describe the chemical potentials of different particles.

It is doubtless that the chemical potentials of quarks in high energy collisions are an important subject. We are interested in measurements of the chemical potentials of quarks. Because of the complex instruments in high energy collisions, we pin our hope on experimental collaborations who work at experimental studies at the LHC, RHIC, and other accelerators. In this paper, we propose two methods which can extract chemical potentials of quarks based on the transverse momentum spectra of different particles. Following our proposals, the experimental collaborations can measure the transverse momentum spectra of some particles and obtain chemical potentials of different quarks.

## 2. The Two Methods

According to the predictions of a statistical model based on chemical and thermal equilibrium at the quark level, a comparison of the measured anti-particle to particle ratios is given by [13]

$$\frac{\rho(\bar{p})}{\rho(p)} = \exp\left(-\frac{6\mu_{u,d}}{T}\right) \tag{1}$$

and

$$\frac{\rho(K^-)}{\rho(K^+)} = \exp\left[-\frac{2(\mu_{u,d} - \mu_s)}{T}\right] = \exp\left(\frac{2\mu_s}{T}\right)\left[\frac{\rho(\bar{p})}{\rho(p)}\right]^{1/3}, \tag{2}$$

where $\rho$ denotes the number density of related particles, $T$ denotes the temperature of interacting system at the stage of chemical freeze-out, $\mu_{u,d}$ is the chemical potential of up quark or down quark due to that the two chemical potentials are nearly the same, and $\mu_s$ is the chemical potential of strange quark. One can see that the chemical potentials of up (down) and strange quarks can be extracted from the ratios of anti-particle yield to particle yield. In the extraction, we have to know the temperature of the interacting system at the stage of chemical freeze-out.

Generally, by fitting the available experimental data, the values of chemical potentials for the up (down) and strange quarks can be obtained from the ratios of $\bar{p}/p$, $K^-/K^+$, and $\pi^-/\pi^+$ in a wide transverse momentum range [14]. If we differentiate the chemical potentials of up and down quarks, and use different temperatures of the interacting system at the stage of kinetic freeze-out for the productions of different particles, we are hopeful to study the chemical potentials of quarks in details [15]. Particularly, we hope to extract the chemical potentials of all six flavors of quarks. However, because of the very large mass and very short lifetime of the top quark [16], it behaves differently from other quarks. In fact, the top quark decays before it hadronizes, which provides an independent opportunity for observations of the top quark. Consequently we don't discuss the top quark in this paper.

The first method is based on Eqs. (1) and (2) which are discussed in reference [13]. Let $\mu_i$ denote the chemical potentials of the $i$-th flavor quark, where $i = u, d, s, c,$ and $b$ denote the up, down, strange, charm, and bottom quarks, respectively. In principle, we can use the ratios of different negatively/positively particles to give relations among different $\mu_i$. The values of $\mu_i$ are expected from these relations. In Table 1, we list some hadrons and their quark structures [17], which can be used in our analyses. A few appropriate combinations can be used to extract $\mu_i$. As an example，we choose $\bar{p}$, $p$, $K^-$, $K^+$, $\pi^-$, $\pi^+$, $D^-$, $D^+$, $B^-$ and $B^+$ from the table. Then, the ratios are

$$k_p \equiv \frac{\rho(\bar{p})}{\rho(p)} = \exp\left(-\frac{(2\mu_u + \mu_d)}{T_p}\right) \Big/ \exp\left(\frac{(2\mu_u + \mu_d)}{T_p}\right) = \exp\left(-\frac{2(2\mu_u + \mu_d)}{T_p}\right), \tag{3}$$

$$k_K \equiv \frac{\rho(K^-)}{\rho(K^+)} = \exp\left(-\frac{(\mu_u - \mu_s)}{T_K}\right) \Big/ \exp\left(\frac{(\mu_u - \mu_s)}{T_K}\right) = \exp\left(-\frac{2(\mu_u - \mu_s)}{T_K}\right), \tag{4}$$

$$k_\pi \equiv \frac{\rho(\pi^-)}{\rho(\pi^+)} = \exp\left(-\frac{(\mu_u - \mu_d)}{T_\pi}\right) \Big/ \exp\left(\frac{(\mu_u - \mu_d)}{T_\pi}\right) = \exp\left(-\frac{2(\mu_u - \mu_d)}{T_\pi}\right), \tag{5}$$

$$k_D \equiv \frac{\rho(D^-)}{\rho(D^+)} = \exp\left(-\frac{(\mu_c - \mu_d)}{T_D}\right) \Big/ \exp\left(\frac{(\mu_c - \mu_d)}{T_D}\right) = \exp\left(-\frac{2(\mu_c - \mu_d)}{T_D}\right), \tag{6}$$

$$k_B \equiv \frac{\rho(B^-)}{\rho(B^+)} = \exp\left(-\frac{(\mu_u - \mu_b)}{T_B}\right) \Big/ \exp\left(\frac{(\mu_u - \mu_b)}{T_B}\right) = \exp\left(-\frac{2(\mu_u - \mu_b)}{T_B}\right), \tag{7}$$

where $k_j$ ($j = p, K, \pi, D, B$) and $T_j$ denote the ratios of related negatively/positively charged particles and effective temperatures obtained from the transverse momentum spectra of related particles, respectively. Because of temperatures being extracted from the transverse momentum spectra, we discuss our results at the stage of kinetic freeze-out instead of chemical freeze-out. At the same time, we use approximately the effective temperatures instead of the freeze-out temperature.

We can use different distributions to describe the transverse momentum spectra and to obtain the ratios and temperatures. These distributions include but not limited to the Boltzmann distribution, Fermi-Dirac/Bose-Einstein distribution, Tsallis distribution, and others. In these distributions, temperatures are related to chemical potentials of hadrons. We can obtain some kinetic freeze-out curves from these distributions based on collisions at different energies. Particularly, for the Tsallis distribution, temperature is dependent on chemical potential of hadron, and non-extensive parameter determines the strength of such dependence. In the descriptions of transverse momentum spectra, the Tsallis distribution uses three free parameters in contrast to the standard (Boltzmann, Fermi-Dirac, and Bose-Einstein) distributions which use only two free parameters.

The related chemical potentials of quarks are

$$\mu_u = -\frac{1}{6}\left(T_p \ln k_p + T_\pi \ln k_\pi\right), \tag{8}$$

$$\mu_d = -\frac{1}{6}\left(T_p \ln k_p - 2T_\pi \ln k_\pi\right), \tag{9}$$

$$\mu_s = -\frac{1}{6}\left(T_p \ln k_p + T_\pi \ln k_\pi - 3T_K \ln k_K\right), \tag{10}$$

$$\mu_c = -\frac{1}{6}\left(T_p \ln k_p - 2T_\pi \ln k_\pi + 3T_D \ln k_D\right), \tag{11}$$

$$\mu_b = -\frac{1}{6}\left(T_p \ln k_p + T_\pi \ln k_\pi - 3T_B \ln k_B\right). \tag{12}$$

Experimentally, the ratios $k_p$, $k_K$, and $k_\pi$ are usually measured, and the values of $\mu_u$, $\mu_d$, and $\mu_s$ can be obtained [14, 15]. If the ratios $k_D$ and $k_B$ are measured in experiments, the values of $\mu_c$ and $\mu_b$ can also be obtained. To extract different $\mu_i$, we propose experimental collaborations to measure simultaneously not only the transverse momentum spectra of $\bar{p}$, $p$, $K^-$, $K^+$, $\pi^-$, and $\pi^+$, but also those of $D^-$, $D^+$, $B^-$ and $B^+$ in high energy collisions.

The second method is based on the chemical potentials of related hadrons. We can choose different combinations of related hadrons from Table 1. As an example, we choose $\Delta^{++}$, $\Delta^-$, $\Omega^-$, $D^+$, and $B^+$. Different distributions can be used to describe the transverse momentum spectra. The standard distributions for particles produced in a rest source can be written as

$$f_{p_T}(p_T) = \frac{1}{N}\frac{dN}{dp_T} = Cp_T \left[\exp\left(\frac{\sqrt{p_T^2 + m_0^2} - \mu}{T_B}\right) \pm 1\right]^{-1}, \tag{13}$$

where $N$ is the number of particles, $C$ is the normalization constant of the probability distribution, $T_B$ is the effective temperature parameter of the interacting system measured at the stage of kinetic freeze-out, $m_0$ is the rest mass of the considered particle, $\mu$ is the chemical potential of the considered particle, $+1$ denotes fermions and the equation is the Fermi-Dirac distribution, and $-1$ denotes bosons and the equation is the Bose-Einstein distribution. If we neglect $\pm 1$ in the above equation, it will reduce to the Boltzmann distribution. Meanwhile, if we neglect $\mu$ in the above equation, it will reduce to the simplest Boltzmann distribution.

In some cases, the transverse momentum spectra cannot be described by a single standard distribution, but a two- or three-component standard distribution which reflects temperature fluctuations in the interacting system and can be uniformly described by the Tsallis distribution. This is the result of the multisource thermal model [18–20]. In the case of using a non-single standard distribution, we can obtain the temperature and chemical potential by a weighted average respectively. After getting the chemical potentials $\mu_{\Delta^{++}}$, $\mu_{\Delta^-}$, $\mu_{\Omega^-}$, $\mu_{D^+}$, and $\mu_{B^+}$ which correspond to those of $\Delta^{++}$, $\Delta^-$, $\Omega^-$, $D^+$, and $B^+$, respectively, we have the following chemical potentials of quarks based on the quark structures of the mentioned particles

$$\mu_u = \frac{1}{3}\mu_{\Delta^{++}}, \tag{14}$$

$$\mu_d = \frac{1}{3}\mu_{\Delta^-}, \tag{15}$$

$$\mu_s = \frac{1}{3}\mu_{\Omega^-}, \tag{16}$$

$$\mu_c = \mu_d + \mu_{D^+} = \frac{1}{3}\mu_{\Delta^-} + \mu_{D^+}, \tag{17}$$

$$\mu_b = \mu_u - \mu_{B^+} = \frac{1}{3}\mu_{\Delta^{++}} - \mu_{B^+}. \tag{18}$$

To obtain the chemical potentials of quarks, we would like to propose the experimental collaborations to measure simultaneously the transverse momentum spectra of $\Delta^{++}$, $\Delta^-$, $\Omega^-$, $D^+$, and $B^+$ in high energy collisions.

As an alterable treatment of the second method, to obtain $\mu_u$, $\mu_d$, and $\mu_s$, we can also analyze the transverse momentum spectra of $\bar{p}$, $p$, $K^-$, $K^+$, $\pi^-$, and $\pi^+$ by using the Fermi-Dirac distribution. In the case of analyzing $p$, $K^+$, and $\pi^+$, according to the quark structures of these hadrons, we have

$$\mu_u - \mu_d = \mu_{\pi^+}, \tag{19}$$

$$\mu_u - \mu_s = \mu_{K^+}, \tag{20}$$

$$2\mu_u + \mu_d = \mu_p, \tag{21}$$

where $\mu_{\pi^+}$, $\mu_{K^+}$, and $\mu_p$ denote the chemical potentials of $\pi^+$, $K^+$, and $p$, respectively.

Thus, we have

$$\mu_u = \frac{1}{3}\left(\mu_p + \mu_{\pi^+}\right), \tag{22}$$

$$\mu_d = \frac{1}{3}\left(\mu_p - 2\mu_{\pi^+}\right), \tag{23}$$

$$\mu_s = \frac{1}{3}\left(\mu_p + \mu_{\pi^+}\right) - \mu_{K^+}. \tag{24}$$

Further,

$$\mu_c = \frac{1}{3}\left(\mu_p - 2\mu_{\pi^+}\right) + \mu_{D^+}, \tag{25}$$

$$\mu_b = \frac{1}{3}\left(\mu_p + \mu_{\pi^+}\right) - \mu_{B^+}. \tag{26}$$

**3. Applications**

We studied chemical potentials of up, down, and strange quarks at RHIC and LHC energies in our previous work [14, 15] by using the first method. To avoid repetition in the present work, we analyze the transverse momentum spectra of $\bar{p}$, $p$, $K^-$, $K^+$, $\pi^-$, and $\pi^+$ produced in Pb-Pb collisions at the Super Proton Synchrotron (SPS), and use different distributions to describe the spectra. In the first method, we use the simplest two-component Boltzmann distribution

$$f_{p_T}(p_T) = K_1 C_1 p_T \exp\left(-\frac{\sqrt{p_T^2 + m_0^2}}{T_1}\right) + (1 - K_1) C_2 p_T \exp\left(-\frac{\sqrt{p_T^2 + m_0^2}}{T_2}\right), \tag{27}$$

where $K_1$ denotes the contribution ratio of the first component, $T_1$ and $T_2$ denote the measured effective temperatures of the first and second components respectively, and $C_1$ and $C_2$ denote the normalization constants of the first and second components respectively. Because the first method emphasizes the ratios of negatively/positively charged particles, we have neglected the chemical potential of particles in the Boltzmann distribution and used the simplest Boltzmann distribution. In the second method, we use the two-component Fermi-Dirac distribution

$$f_{p_T}(p_T) = K_1 C_1 p_T \left[\exp\left(\frac{\sqrt{p_T^2 + m_0^2} - \mu_1}{T_1}\right) + 1\right]^{-1}$$

$$+ (1 - K_1) C_2 p_T \left[\exp\left(\frac{\sqrt{p_T^2 + m_0^2} - \mu_2}{T_2}\right) + 1\right]^{-1}. \tag{28}$$

Because the second method emphasizes the chemical potential of charged particles, we have saved this quantity in the Fermi-Dirac distribution.

Figure 1 presents the transverse momentum spectra of (a,c,e) $\pi^+$ (squares), $K^+$ (circles), and $p$ (triangles), as well as (b,d,f) $\pi^-$ (squares), $K^-$ (circles), and $\bar{p}$ (triangles) produced in Pb-Pb collisions at center-of-mass energy per nucleon pair of 17.3 GeV (one of the SPS energies). Figures 1(a)/1(b), 1(c)/1(d), and 1(e)/1(f) correspond to the centralities of 0–5%, 12.5–23.5%, and 33.5–80%, respectively. The symbols represent the experimental data of the NA49 Collaboration [21], and the solid and dashed curves are the results of Eqs. (27) and (28) respectively. The corresponding parameter values with $\chi^2$ per degree of freedom ($\chi^2$/dof) are listed in Tables 2 and 3 respectively, where the parameter values are obtained by using the method of least-square-fit, and the values for positively/negatively charged particles are given in terms of "value$_1$/value$_2$" in the case of the two values being different. We can see that the two distributions describe the experimental spectra, and the two calculated results are almost the same due to proper choice of parameters.

According to the first method [Eqs. (8) –(10)] and the solid curves in Figure 1, the calculated chemical potentials of up, down, and strange quarks in Pb-Pb collisions at 17.3 GeV are displayed in Figures 2(a), 2(b), and 2(c), respectively, where the solid, dashed, and dotted curves correspond to the centralities of 0–5%, 12.5–23.5%, and 33.5–80%, respectively. In the case of the temperatures for positively and negatively charged particles being different, we have used the mean temperature. Similarly, according to the first method and the dashed curves in Figure 1, the calculated dependences of $\mu_i$ on $p_T$ are shown in Figure 3. According to Figures 2 and 3, the mean values of $\mu_i$ are listed in Table 4. One can see that Figures 2 and 3 are similar to each other. The simplest two-component Boltzmann distribution and the two-component Fermi-Dirac distribution result in similar chemical potentials of quarks. We would like to point out that, although Figures 2 and 3 show the dependences of chemical potentials of quarks on transverse momentum, it does not mean that there is a correlation between $\mu_i$ and $p_T$. At the same time, it does not mean really different chemical potentials for different phase spaces. In our opinion, the situations presented in Figures 2 and 3 reflect statistical fluctuations in experimental measurements of transverse momentum spectra. Rather, as presented in Table 4, are more physically.

According to chemical potential values in Table 3 and the second method [Eqs. (22) –(24)], we obtain $\mu_u \approx 170$ MeV, $\mu_d \approx 160$ MeV, and $\mu_s \approx 100$ MeV for three centralities in Pb-Pb collisions at 17.3 GeV. Qualitatively, $\mu_u \approx \mu_d > \mu_s$, which is consistent with values in Table 4, though both the quantities are different. Because the transverse momentum spectra are not sensitive to the chemical potentials of hadrons in Eq. (28), the second method is not a refined method. Then, the chemical potentials of quarks obtained from the second method are not sensitive on the parameterization of transverse momentum spectra. Contrarily, the first method is based on the ratios of negatively/positively charged particles, which is sensitive to the transverse momentum spectra and renders the first method being a relative refined method. This means that the chemical potentials of quarks obtained from the first method are sensitive on the parameterization of transverse momentum spectra. In addition, the definition and extraction of baryon chemical potentials at kinetic freeze-out in experiments imply that the second method is a deep-set method comparing with the first one.

4. **Conclusions and Discussions**

The comprehension of the quark chemical potentials is very important in studying the phase transition or chemical and thermal equilibriums. In this paper, we have presented two methods which can be used to extract the chemical potentials of quarks in high energy particle-particle, particle-nucleus, and nucleus-nucleus collisions. Because the chemical potentials of up, down, and strange quarks in nuclear collisions at RHIC and LHC energies were studied in our previous work [14, 15], we focus our attention on this subject at the lower SPS energy. Our previous and present studies can only give chemical potentials of some quarks due to limited available experimental data.

The first method is based on the ratios of negatively/positively charged particles which includes $\bar{p}/p$, $K^-/K^+$, $\pi^-/\pi^+$, $D^-/D^+$, and $B^-/B^+$. The temperatures extracted from the transverse momentum spectra of the related hadrons are also needed. Eqs. (8)–(12) are the expressions of quark chemical potentials on the first method. The second method is based on the chemical potentials of some particles such as $\Delta^{++}$, $\Delta^{-}$, $\Omega^{-}$, $D^{+}$, and $B^{+}$; or the same particles as those used in the first method. To obtain the particle chemical potentials, we need to describe the transverse momentum spectra of the related hadrons. In the descriptions of the hadron transverse momentum spectra, we can obtain the chemical potentials of the related hadrons. Eqs. (14)–(18) or Eqs. (22)–(26) are the expressions of quark chemical potentials on the second method.

To extract the quark chemical potentials, we would like to propose the experimental collaborations to measure simultaneously not only the transverse momentum spectra of $\bar{p}$, $p$, $K^{-}$, $K^{+}$, $\pi^{-}$, and $\pi^{+}$, but also those of $D^{-}$, $D^{+}$, $B^{-}$, and $B^{+}$ (even those of $\Delta^{++}$, $\Delta^{-}$, and $\Omega^{-}$) in nuclear collisions at RHIC and LHC energies. In other words, we need the transverse momentum spectra of these hadrons in the same experimental condition. Because the quark structures listed in Table 1 can give similar combinations by different sets of hadrons, the hadrons proposed in the above can be replaced by others. Meanwhile, the two methods can be verified each other in a real extraction.

As an example, the present results from SPS energy show that $\mu_u \approx \mu_d \approx 110 - 150$ MeV and $\mu_s \approx 30 - 60$ MeV from the first method; and $\mu_u \approx \mu_d \approx 160 - 170$ MeV and $\mu_s \approx 100$ MeV from the second method. In the first method, the rations of negatively/positively charged particles are sensitive to the transverse momentum spectra. In the second method, the chemical potentials in Eq. (28) are not sensitive to the transverse momentum spectra. Therefore, we think that the first method is a relative refined method comparing with the second one. At the same time, the chemical potentials of quarks obtained from the first method are sensitive on the parameterization of transverse momentum spectra, in contrarily to those obtained from the second method which are not sensitive on the parameterization.

In the above discussions, since the chemical potentials of quarks at RHIC and LHC energies were studied in our previous work [14, 15], we essentially address lower SPS energy in the present work. Generally, there is no well and unique defined quark chemical potential since trajectories in the temperature-chemical potential ($T - \mu$) plane, e.g. for constant entropy, do not allow to talk about a unique quark chemical potential. The way usually taken is to define the baryon chemical potential at freeze-out of the hadrons which is experimentally well established. This implies that the second method is a deep-set method comparing with the first one.

Usually, the transverse momentum spectrum does not provide direct information on the thermal temperature due to radial flow. Because of the effect of radial flow, one can obtain a wider transverse momentum spectrum comparing with a purely thermal emission process. The measured effective temperatures extracted from the present work are in fact larger than the real situations, i.e. the present work overestimates the temperatures. Considering the effect of radial flow, we should revise the effective temperatures obtained in the present work by subtracting a small amount which affects slightly the chemical potentials of quarks. On the other side, the two-component model seems to be very limited in applications. A revision of radial flow (blast-wave) can reduce possibly the two-component model to the standard distributions. The related blast-wave revision of the standard distributions by using the Monte Carlo method is ongoing by us.

Comparing with our previous work [14, 15], the progress of the present work is obvious. Firstly, our previous work used only one method, and the present work has used two methods and given some comparisons for the two methods. Secondly, our previous work analyzed only the chemical potentials of up, down, and strange quarks, and the present work has analyzed not only the chemical potentials of up, down, and strange quarks, but also the possible chemical potentials of charm and bottom quarks. Thirdly, our previous work dealt with RHIC and LHC energies which result in low chemical potentials which can be approximately neglected, and the present work has dealt with SPS energy which results in high chemical potentials which are in fact considerable.

High energy collisions are very complex, and contain abundant information such as flow effects [22], phase transition [23], entropy density [24], cold nuclear effect, nuclear stopping power, long range correlation, short range correlation, soft excitation process, hard scattering process, and so forth. These rich and colorful phenomena are related to experimental data which include but not limited to azimuthal distributions and correlations, rapidity distributions and correlations, transverse energy distributions, transverse momentum spectra, and others. The present work can supply a referenced worthiness for the study of transverse momentum spectra. As a byproduct in the study, one can obtain the chemical potentials of quarks, which is useful for the study of phase diagram.

**Conflict of Interests**

The authors declare that there is no conflict of interests regarding the publication of this paper.

**Acknowledgments**

This work was supported by the National Natural Science Foundation of China (under Grant No. 10975095), the Open Research Subject of the Chinese Academy of Sciences Large-Scale Scientific Facility (under Grant No. 2060205), the Shanxi Provincial Natural Science Foundation (under Grant No. 2013021006), and the Foundation of Shanxi Scholarship Council of China (under Grant No. 2012-012).

Table 1. The quark structures of some hadrons [17].

| Type | Hadron | Quark structure | Hadron | Quark structure | Hadron | Quark structure | Hadron | Quark structure |
|---|---|---|---|---|---|---|---|---|
| Meson | $\pi^+$ | $u\bar{d}$ | $K^-$ | $s\bar{u}$ | $\bar{D}^0$ | $u\bar{c}$ | $\bar{B}^0$ | $b\bar{d}$ |
| | $\pi^-$ | $d\bar{u}$ | $K^0$ | $d\bar{s}$ | $D_s^+$ | $c\bar{s}$ | $B_s^0$ | $s\bar{b}$ |
| | $\pi^0$ | $(u\bar{u}-d\bar{d})/\sqrt{2}$ | $\bar{K}^0$ | $s\bar{d}$ | $D_s^-$ | $s\bar{c}$ | $\bar{B}_s^0$ | $b\bar{s}$ |
| | $\rho^+$ | $u\bar{d}$ | $D^+$ | $c\bar{d}$ | $B^+$ | $u\bar{b}$ | $B_c^+$ | $c\bar{b}$ |
| | $\rho^-$ | $d\bar{u}$ | $D^-$ | $d\bar{c}$ | $B^-$ | $b\bar{u}$ | $B_c^-$ | $b\bar{c}$ |
| | $K^+$ | $u\bar{s}$ | $D^0$ | $c\bar{u}$ | $B^0$ | $d\bar{b}$ | | |
| Baryon | $p$ | $uud$ | $\Lambda_c^+$ | $udc$ | $\Sigma_c^0$ | $ddc$ | $\Xi_{cc}^{++}$ | $ucc$ |
| | $n$ | $udd$ | $\Lambda_b^0$ | $udb$ | $\Xi^0$ | $uss$ | $\Xi_{cc}^+$ | $dcc$ |
| | $\Delta^{++}$ | $uuu$ | $\Sigma^+$ | $uus$ | $\Xi^-$ | $dss$ | $\Omega_c^0$ | $ssc$ |
| | $\Delta^+$ | $uud$ | $\Sigma^-$ | $dds$ | $\Xi_c^+$ | $usc$ | $\Omega^-$ | $sss$ |
| | $\Delta^-$ | $ddd$ | $\Sigma^0$ | $uds$ | $\Xi_c^0$ | $dsc$ | $\Omega_b^-$ | $ssb$ |
| | $\Delta^0$ | $udd$ | $\Sigma_c^{++}$ | $uuc$ | $\Xi_b^0$ | $usb$ | $\Omega_{cc}^+$ | $scc$ |
| | $\Lambda^0$ | $uds$ | $\Sigma_c^+$ | $udc$ | $\Xi_b^-$ | $dsb$ | | |

Table 2. Parameter values corresponding to the solid curves in Figure 1. The relative errors for the effective temperatures and ratios are about 5% and 3% respectively. The values for positively/negatively charged particles are given in terms of "value$_1$/value$_2$" in the case of the two values being different.

| Centrality | Particle | $T_1$ (GeV) | $k_1$ | $T_2$ (GeV) | $\chi^2$/dof |
|---|---|---|---|---|---|
| 0-5% | $\pi^+/\pi^-$ | 0.125/0.183 | 0.907/0.963 | 0.250/0.296 | 0.659/0.728 |
| | $K^+/K^-$ | 0.172/0.178 | 0.833/0.887 | 0.278/0.275 | 0.755/0.672 |
| | $p/\bar{p}$ | 0.274/0.262 | 0.999 | 0.100 | 1.649/1.650 |
| 12.5-23.5% | $\pi^+/\pi^-$ | 0.160/0.174 | 0.903/0.924 | 0.262/0.274 | 1.377/0.738 |
| | $K^+/K^-$ | 0.167/0.177 | 0.789 | 0.274/0.257 | 0.844/0.938 |
| | $p/\bar{p}$ | 0.274/0.250 | 0.999 | 0.100 | 1.493/1.844 |
| 33.5-80% | $\pi^+/\pi^-$ | 0.162/0.174 | 0.786/0.933 | 0.240/0.272 | 0.804/0.579 |
| | $K^+/K^-$ | 0.167/0.177 | 0.833/0.857 | 0.264/0.257 | 0.365/0.463 |
| | $p/\bar{p}$ | 0.265/0.242 | 0.999 | 0.100 | 1.065/1.224 |

Table 3. Parameter values corresponding to the dashed curves in Figure 1. The relative errors for the effective temperatures, chemical potentials, and ratios are about 5%, 10%, and 3%, respectively. The values for positively/negatively charged particles are given in terms of "value$_1$/value$_2$" in the case of the two values being different.

| Centrality | Particle | $T_1$ (GeV) | $\mu_1$ (MeV) | $k_1$ | $T_2$ (GeV) | $\mu_2$ (MeV) | $\chi^2$/dof |
|---|---|---|---|---|---|---|---|
| 0-5% | $\pi^+/\pi^-$ | 0.140/0.160 | 9/-9.5 | 0.795/0.932 | 0.247/0.275 | 11/-12 | 0.890/0.718 |
| | $K^+/K^-$ | 0.190 | 70/-70 | 0.724/0.894 | 0.277 | 70/-70 | 0.504/0.311 |
| | $p/\bar{p}$ | 0.273/0.271 | 500/-500 | 0.999 | 0.100 | 500/-500 | 0.744/1.522 |
| 12.5-23.5% | $\pi^+/\pi^-$ | 0.140/0.150 | 9/-9.5 | 0.823/0.901 | 0.247/0.259 | 11/-12 | 0.890/0.273 |
| | $K^+/K^-$ | 0.190 | 70/-70 | 0.765/0.886 | 0.277/0.270 | 70/-70 | 0.504/0.517 |
| | $p/\bar{p}$ | 0.273/0.257 | 500/-500 | 0.999 | 0.100 | 500/-500 | 0.744/1.258 |
| 33.5-80% | $\pi^+/\pi^-$ | 0.150/0.155 | 9/-9.5 | 0.853/0.914 | 0.247/0.260 | 11/-12 | 0.799/0.264 |
| | $K^+/K^-$ | 0.190 | 70/-70 | 0.865/0.913 | 0.277/0.270 | 70/-70 | 0.256/0.290 |
| | $p/\bar{p}$ | 0.255/0.248 | 500/-500 | 0.999 | 0.100 | 500/-500 | 0.530/0.810 |

Table 4. Mean values of quark chemical potentials obtained from Figures 2 and 3.

| Figure | Centrality | $\bar{\mu}_u$ (MeV) | $\bar{\mu}_d$ (MeV) | $\bar{\mu}_s$ (MeV) |
|---|---|---|---|---|
| 2 | 0-5% | $135.029 \pm 0.285$ | $145.063 \pm 1.517$ | $59.152 \pm 0.818$ |
| | 12.5-23.5% | $135.949 \pm 0.720$ | $145.236 \pm 1.201$ | $43.807 \pm 1.002$ |
| | 33.5-80% | $111.674 \pm 0.381$ | $142.207 \pm 2.002$ | $31.975 \pm 0.686$ |
| 3 | 0-5% | $130.158 \pm 0.392$ | $143.258 \pm 0.966$ | $38.997 \pm 1.202$ |
| | 12.5-23.5% | $137.910 \pm 0.430$ | $142.949 \pm 0.931$ | $36.969 \pm 0.889$ |
| | 33.5-80% | $108.595 \pm 0.109$ | $121.513 \pm 0.716$ | $27.699 \pm 0.909$ |

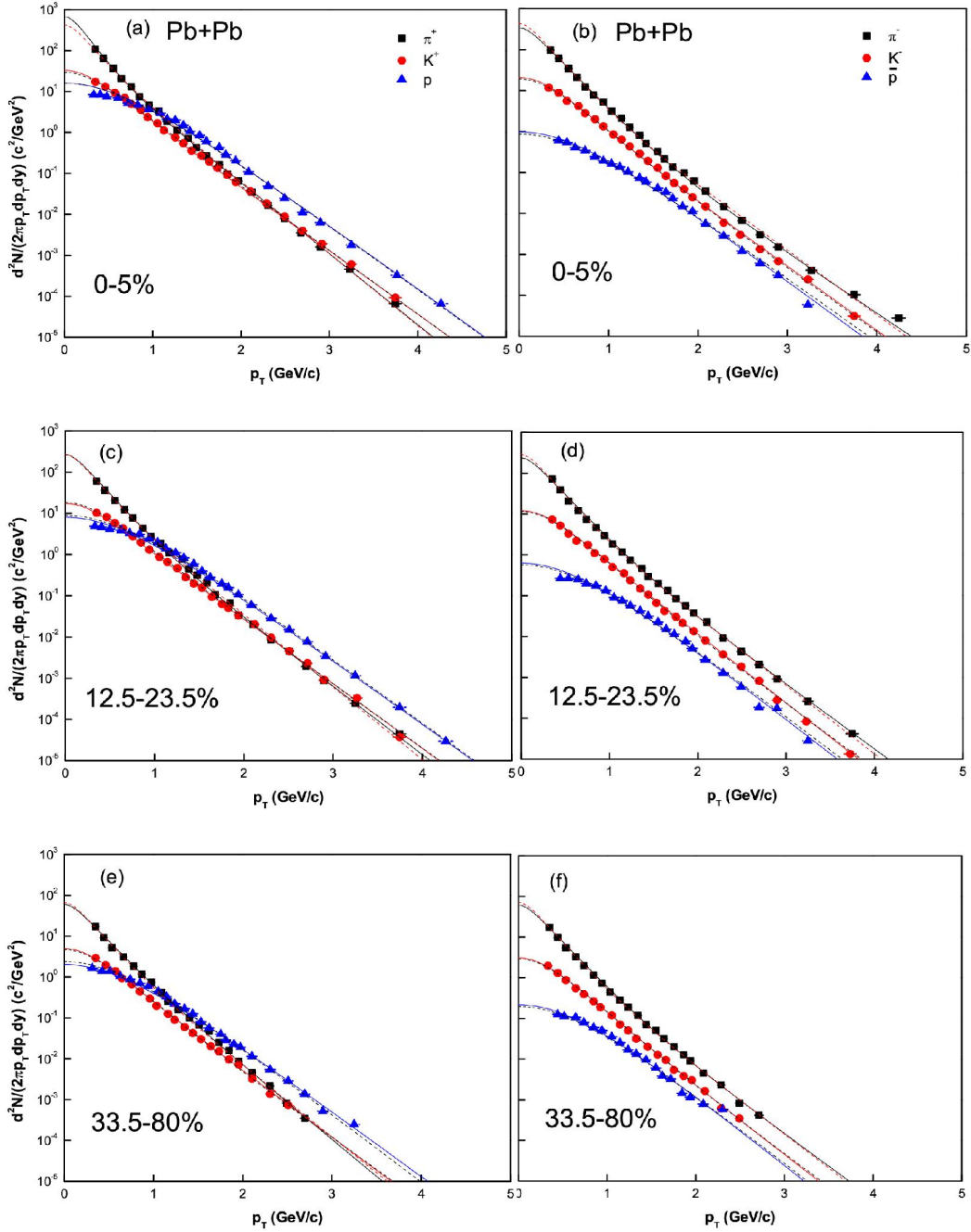

Figure 1. Transverse momentum spectra of (a,c,e) $\pi^+$ (squares), $K^+$ (circles), and $p$ (triangles), as well as (b,d,f) $\pi^-$ (squares), $K^-$ (circles), and $\bar{p}$ (triangles) produced in Pb-Pb collisions at 17.3 GeV. Figures 1(a)/1(b), 1(c)/1(d), and 1(e)/1(f) correspond to the centralities of 0–5%, 12.5–23.5%, and 33.5–80%, respectively. The symbols represent the experimental data of the NA49 Collaboration [21], and the solid and dashed curves are the results of Eqs. (27) and (28) respectively.

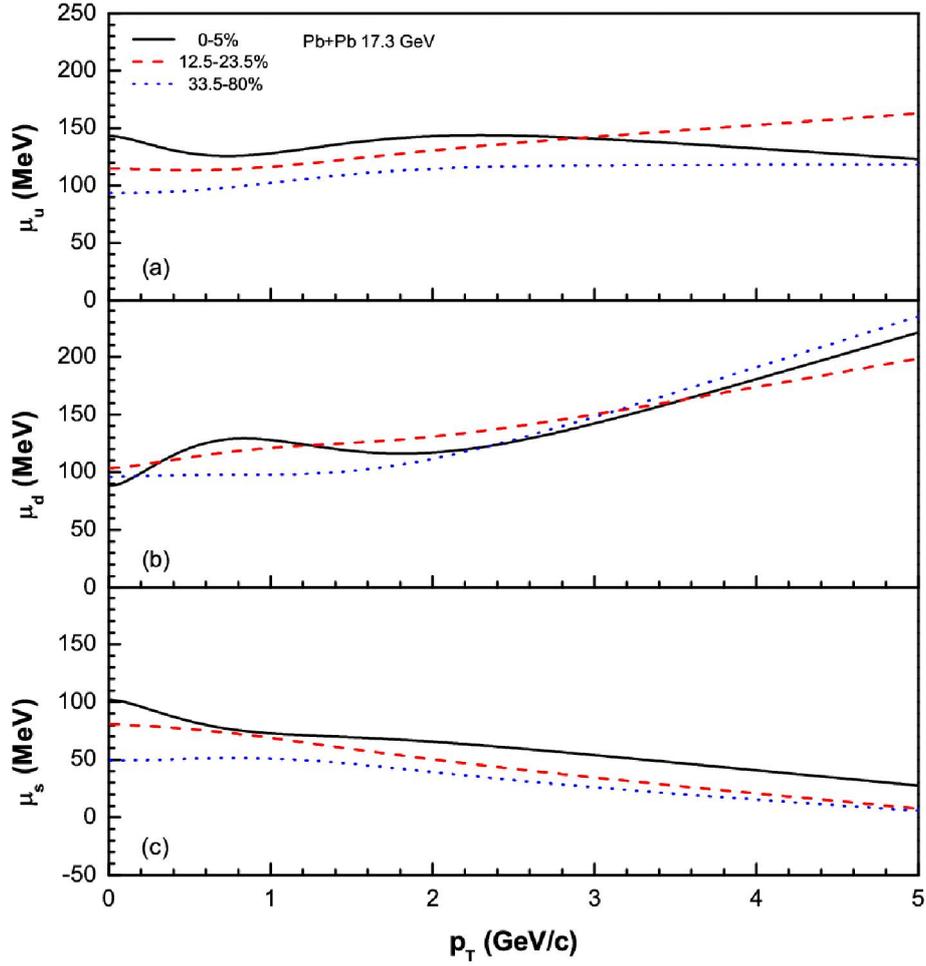

Figure 2. According to the solid curves in Figure 1, the calculated chemical potentials of (a) up, (b) down, and (c) strange quarks in Pb-Pb collisions at 17.3 GeV are displayed, where the solid, dashed, and dotted curves correspond to the centralities of 0–5%, 12.5–23.5%, and 33.5–80%, respectively.

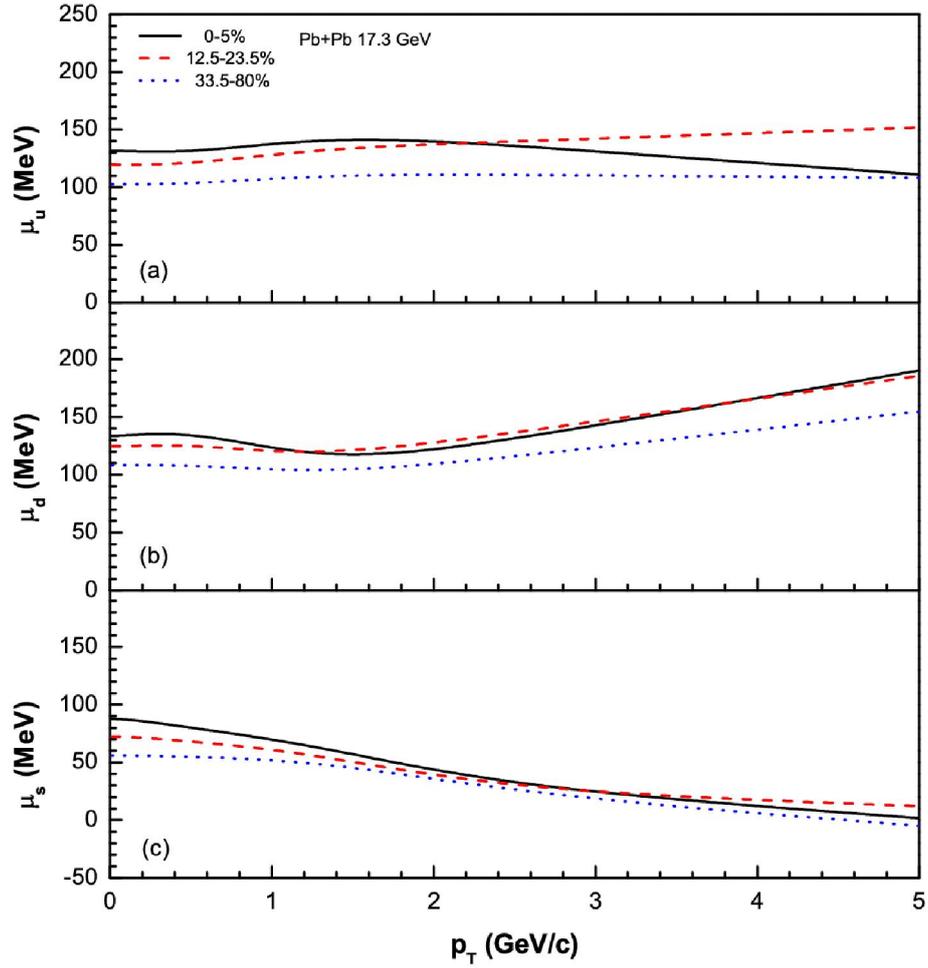

Figure 3. The same as that for Figure 2, but showing the results according to the dashed curves in Figure 1.